\documentclass[12pt]{article}
\usepackage{amsmath}
\usepackage{amssymb,amsthm}
\usepackage{color}

\numberwithin{equation}{section}

\theoremstyle{definition}

\theoremstyle{remark}

\title{The Meissner effect in the ground state of free charged Bosons in a constant magnetic field}
\author{Walter F. Wreszinski\\
        Instituto de Fisica USP\\
        Rua do Mat\~{a}o, s.n., Travessa R 187\\
        05508-090 S\~{a}o Paulo, Brazil\\
        \texttt{wreszins@gmail.com}
        }
\begin{document}
\maketitle
\begin{abstract}
The model of free charged Bosons in an external constant magnetic field inside a cylinder, one of the few locally gauge covariant systems amenable to analytic treatment, is rigorously investigated in the semiclassical approximation. The model was first studied by Schafroth and is suitable for the description of quasi-bound electron pairs localized in physical space, so-called Schafroth pairs, which occur in certain compounds. Under the assumption of existence of a solution of the semiclassical problem for which the ground state (g.s.) expectation value of the current $<\vec{j}(\vec{x})>$ is of the London form, i.e., $<\vec{j}(\vec{x})> = -c |\phi_{0}(\vec{x})|^{2} \vec{A}(\vec{x})$, where c is a positive constant, $\vec{A}$ the vector potential and $\phi_{0}$ the one-particle g.s. wave-function. as well as some regularity assumptions, the magnetic induction may be proved to decay exponentially from its value on the surface of the cylinder. An important role is played by a theorem on the pointwise monotonicity of the ground state wave-function on the potential.     
\end{abstract}

\section{Introduction: motivation and synopsis of the paper}

The Meissner effect is perhaps the prototypical electrodynamic property of superconductors, and is one of the most spectacular effects in physics. It may be stated as follows (see, e.g., \cite{MaRo}, pg. 162):

\textbf{Meissner effect} If a superconductor sample is submitted to a magnetic (applied) field $\vec{H}$ and then cooled to a temperature $T$ below the transition temperature, there is a critical field $H_{c}(T)>0$ such that, if $|\vec{H}| < H_{c}(T)$, the field is expelled from all points of the sample situated sufficiently far from from the surface, i.e., beyond a certain distance from the surface called the ''penetration depth''.

It follows from the above that the Meissner effect is an extreme instance of the magnetic effects found in substances, called diamagnetic, which are repelled from, e.g., a strong electromagnet with a sharply pointed pole piece - one might call them complete diamagnets \cite{Schaf1}. As such, it is necessarily a \textbf{quantum} phenomenon, because classical physics explains neither paramagnetism or diamagnetism in an equilibrium state (thermal or ground state) \cite{FeIII}. The word ''equilibrium'' is important here, and therefore it should be mentioned that, in contrast to the vanishing of electrical resistence, the Meissner effect is a phenomenon in thermal equilibrium - see \cite{Schaf1}, pg. 299 for a full justification of this assertion. An important part of the argument is directly connected to the present paper, and therefore we mention it here: the quantum aspect of the theory, when treated semiclassically, depends only on the expectation value of the current $\vec{j}(\vec{x})$ in the equilibrium state $<.>$ (ground state or thermal), see the forthcoming (2.7), in case the state is an equilibrium state. When no magnetic field is present, this equilibrium state may be assumed to be invariant under the time reversal operator $\tau$. However, $\tau(\vec{j}(\vec{x})) = -\vec{j}(\vec{x})$, hence $<\vec{j}(\vec{x})> = <\tau(\vec{j}(\vec{x}))> = -<\tau(\vec{j}(\vec{x}))> = 0$. This is Bloch's theorem (unpublished), cited in \cite{Schaf1} and \cite{Kad}. Thus, the phenomenon of persistent currents, unlike the Meissner effect, cannot be an equilibrium phenomenon. A generalization of this statement beyond the semiclassical approximation for systems carrying a current was given by Sewell in \cite{Se1}, see also \cite{SeW} for the superfluid case.

The theory of Bardeen, Cooper and Schrieffer \cite{BCS} - see also the textbook accounts, too numerous to cite; we mention only \cite{MaRo}, \cite{Enz} and \cite{Zi}, which will be directly referred to in this paper - has been very successful in explaining a variety of phenomena observed in (ordinary) superconductivity, such as the energy of the superconducting ground state and the energy gap associated with single-particle excitations. An excellent account, to a large extent still up-to-date, is M. R. Schafroth's review, which appeared shortly before his untimely death \cite{Schaf1}, of which we quote: ''The most serious failing of the theory of BCS is, however, its failure to account for the electrodynamical properties of superconductors, such as the Meissner effect and the persistent currents.'' 

Schafroth's remark was due to the fact that the BCS model does not satisfy the requirement of local gauge invariance (or, more properly, local gauge covariance, see section 3): this fact was observed by the authors of \cite{BCS} themselves in a footnote in \cite{BCS}: we shall refer to this problem briefly as the ''gauge problem''. Many very renowned scientists worked on the gauge problem (\cite{An3},\cite{An4},\cite{An5},\cite{Nam}), see also \cite{Sewell}, and further references given there. Today, the best accepted justification seems to be the one by Anderson (\cite{An3},\cite{An4},\cite{An5}), quoted by Kadanoff \cite{Kad} in his recent historical review: ''oscillations of the gap parameter or, equivalently, of the condensate wave-function, produces extra states of the system, states which rescue the gauge invariance of the BCS theory''. The type of oscillation referred to by Kadanoff are the plasma oscillations, whose inclusion should, according to Anderson, favor the particular gauge adopted by BCS \cite{BCS} to study the electrodynamical properties of the ground state. However, as remarked by Schafroth (\cite{Schaf1}, pg. 471), since Anderson's starting point (\cite{An3}, \cite{An4}) was the truncated Hamiltonian of the BCS theory, ''whose relation to the complete Hamiltonian of the metal electrons (which would include the plasma oscillations) is not well-defined, his considerations can be, at most, of heuristic value''. The more systematic construction of the ''collective excited states'' by Bogoliubov \cite{Bog}, with energies lying in the energy gap of single-particle excitations, is spoiled if one attempts to include the Coulomb interaction (\cite{Bog}, \cite{An3}, \cite{An4}), see also the remarks in \cite{Schaf1}, pp 487-488), showing that Anderson's justification of local gauge invariance of the BCS model displays the same kind of instability. Moreover, the validity of the concept of ''approximately gauge invariant'' used by him (\cite{An3},\cite{An4}) is difficult to assess, and the random-phase approximation \cite{An5} ( see also \cite{MaRo} for a particularly lucid discussion) does not provide a way of estimating the errors involved in the approximation. It is certainly impossible to reconcile such approximate arguments with the fact that the BCS model is regarded as one of the few rigorous (mean-field) models in statistical mechanics (see, e.g., \cite{LvH} and references given there).

Although the basic (gauge covariant) model for superconductivity, the H. Froehlich electron-phonon system (\cite{Schaf1}), has so far resisted analysis, a serious alternative to the BCS theory arose when Yang \cite{Ya} proposed that the property of off-diagonal long-range-order (ODLRO), introduced by Penrose and Onsager \cite{PeO} (see \cite{Sewell}, 9.3.2, for the precise definition), provides a characterization of the superconductive phase (see also chapter 5 of \cite{Leg2}). This proposal was taken up by Sewell (\cite{Sewell}, \cite{Sewel}), who provided a simple and elegant proof of the following fact (here roughly stated: for details, see Prop. 9.3.2, pg 218, of \cite{Sewell} or \cite{Sewel}): when the magnetic induction is uniform and the equilibrium state of the system is covariant under space translations and local gauge transformations and posesses the ODLRO property, the magnetic induction vanishes. Further assumptions related to thermodynamic stability (\cite{Sewell}, pp. 223-224) imply the Meissner effect (Prop. 9.4.1, pg. 255, of \cite{Sewell}).

Sewell's analysis captures the conceptual essence of the Meissner effect, up to one important point: it concerns only the bulk of the material, i.e., infinite systems. This is due to the fact that the definition of ODLRO involves a limit whereby the distance between two points tends to infinity (see, again, \cite{Sewell}, pg. 218. It thereby misses the important phenomenon of the penetration depth (see, e.g., \cite{MaRo}, pg. 165). It is our objective in the present paper to fill this gap, at least partially, by revisiting the model of the ground state (g.s) of free charged Bosons in a constant magnetic field, in the semiclassical approximation. In spite of the simplicity of this model, we believe, with Schafroth, who first studied it (also for nonzero temperature) (\cite{Schaf1}\cite{Schaf3}), that it exhibits (in addition to being locally gauge covariant) several of the essential features of real superconductors.

The subject of the present paper may be considered by many to be well understood, even by those who are aware of the ''gauge problem''. They would argue that the macroscopic ''slippery'' wave function $\Phi(\vec{x},t)$, a function varying appreciably only over distances characteristic to variations of the electromagnetic fields themselves (\cite{Kad}, \cite{Leg1}, \cite{Leg2}), provides a justification of the Meissner effect, as shown in several textbook treatments (see \cite{MaRo}, pg. 165 or \cite{FeIII}, 21-9)). It is, however, the previously mentioned concept of ODLRO which provides a quantum, i.e., from first principles justification of the macroscopic wave-function (see chapter 5 of \cite{Leg2} or \cite{Kad}). If one accepts this, there remains, however, a basic unresolved conceptual problem (not just a question of mathematical rigor!): by the previously mentioned theorem by Sewell (Prop. 9.3.2. of \cite{Sewell}) - a nonperturbative result! - ODLRO implies, together with the other standard assumptions of space translation and local gauge invariance) that $\vec{B} = \vec{0}$. It is, however, essential to consider a $\vec{B} \ne \vec{0}$ (with $|\vec{B}|$ sufficiently small) to start with, to show that the field is repelled! This shows that the assumption of a macroscopic wave function is not, a priori, justified.

In section 2 we introduce the model and formulate our basic assumptions (assumptions A.1, A.2 and B). Section 3 is devoted to local gauge covariance and properties of the current. In section 4 we prove our main results. We leave to section 5 - our comments on some basic physical and mathematical features of the effect, as well as open problems and some brief remarks on the connection with the Higgs mechanism. In section 6 - the conclusion - we discuss what is, in our view, the main contribution of the paper.

For a reader less interested in the mathematics, we leave the more extensive details of the proof of the main theorem (theorem 3) to appendix A.

\section{The model: ground state of a free charged Bose gas in an external constant magnetic field in the semiclassical approximation}

The model we shall revisit was studied by Schafroth \cite{Schaf3}, as one of the very few locally gauge invariant systems possibly related to superconductivity. Today, it may be viewed as a model for Schafroth pairs \cite{Schaf2} which are known to occur in certain compounds (\cite{Noz}, \cite{Leg3}), i.e., quasi-bound electron pairs localized in physical space, i.e., for which the spatial extension of the pair wave function, measured by the coherence length, is small compared with the average distance between pairs. In this case all electrons of the band are paired, and the pairs form a dilute Bose gas (see also the discussion in the book by Enz \cite{Enz}, chapter 4, pg. 180 et seq.). The Hamiltonian may be written ($e=2e_{0}$, $e_{0}$ being the electron charge):
\begin{eqnarray*}
H(\vec{A}) = \frac{\hbar^{2}}{2m} \int_{K} (\nabla + \frac{ie}{\hbar} \vec{A}(\vec{x}))\Psi^{*}(\vec{x}) \cdot\\
\cdot (\nabla - \frac{ie}{\hbar} \vec{A}(\vec{x})) \Psi(\vec{x}) d\vec{x}
\end{eqnarray*}$$\eqno{(2.1)}$$
where $\Psi(\vec{x})$ and $\Psi^{*}(\vec{x}$ are the basic destruction and creation operators on symmetric Fock space ${\cal F}_{s}({\cal H})$, with 
${\cal H} = L^{2}(K)$ the (one-particle) Hilbert space of square integrable wave functions on the cyçlinder $K$, assumed to be of radius $R$ and height $L$, centered at the origin, of volume $V= \pi R^{2} L$. We refer to, e.g., \cite{MaRo} pg. 91 - there the $\Psi$ are denoted by $a$ - for the $\Psi$, which satisfy the canonical commutation relations
$$
[\Psi(\vec{x}), \Psi^{*}(\vec{x}^{'})] = \delta(\vec{x}-\vec{x}^{'})
\eqno{(2.2.1)}
$$
or, in smeared form, with
$$
\Psi(f) = \int d\vec{x} f(\vec{x}) \Psi(\vec{x})
\eqno{(2.2.2)}
$$
$$
[\Psi(f),\Psi^{*}(g)] = (f,g)
\eqno{(2.2.3)}
$$
for $f,g \in {\cal H}$, and $(f,g) \equiv \int d\vec{x} \bar{f}(\vec{x}) g(\vec{x})$ the inner (scalar) product in the Hilbert space ${\cal H}$, $\bar{f}$ denoting the complex conjugate and $*$ the Hermitian conjugate or adjoint. $H(\vec{A})$ is the (self-adjoint) second quantization of a one-particle operator on ${\cal H}$ (\cite{MaRo}, pg 101), with certain boundary conditions. We shall, however, use an extended version of this one-particle operator (see the forthcoming (2.8), (2.9)). The current density operator in this model is \cite{Schaf1}, pg. 416:
$$
\vec{j}(\vec{x}) = \vec{j}_{mom}(\vec{x}) + \vec{j}_{Lon,\vec{B}}(\vec{x})
\eqno{(2.3)}
$$
where
$$
\vec{j}_{mom}(\vec{x}) = -\frac{ie\hbar}{2m}:\Psi^{*}(\vec{x})\nabla \Psi(\vec{x}) - \Psi(\vec{x}) \nabla \Psi(\vec{x}):
\eqno{(2.4.1)}
$$
and
$$
\vec{j}_{Lon,\vec{B}}(\vec{x}) = -\frac{e^{2}}{m} \Psi^{*}(\vec{x}) \Psi(\vec{x}) \vec{A}(\vec{x})
\eqno{(2.4.2)}
$$
Above, :: denotes the Wick or normal product (e.g., \cite{MaRo}, pg 100). since (2.4) is the momentum density operator, we call it the momentum density component of the current; $\vec{j}_{Lon,\vec{B}}$ is the London part (\cite{MaRo}, (5.15), pg. 164 et seq.), about which much will be said later on.

Let $(\vec{e}_{\rho},\vec{e}_{\theta},\vec{e}_{z})$ be a positively-oriented orthonormal basis adapted to cylindrical coordinates. We assume that the magnetic induction $\vec{B}$ is uniform outside the cylinder, i.e., that the system is embedded in $\mathbf{R}^{2} \times [-L/2,L/2]$, and
$$
\vec{B}(\rho,\theta,z) = B \vec{e}_{z} \mbox{ for } \rho \ge R
\eqno{(2.5)}
$$
where $B$ is a positive constant.  The \textbf{semiclassical} (static) model will be \textbf{defined} by (2.5), together with the set of Maxwell equations for the magnetic induction $\vec{B}$ (using the MKS system):
$$
(\nabla \cdot \vec{B})(\vec{x}) = 0
\eqno{(2.6.1)}
$$
$$
(\nabla \times \vec{B})(\vec{x}) = \mu_{0} (\Omega_{\vec{B}}, \vec{j}(\vec{x}) \Omega_{\vec{B}})
\eqno{(2.6.2)}
$$
where
$$
\vec{B}(\vec{x}) = (\nabla \times \vec{A})(\vec{x})
\eqno{(2.6.3)}
$$
and $\Omega_{\vec{B}}$ is the ground state of $H(\vec{A})$, assumed unique (see later). Note that the g.s. $\Omega_{\vec{B}}$ depénds itself on the solution to (2.6). At present no existence theorem for the semiclassical model is known (see also further remarks in section 5), and we shall have to assume:

\textbf{Assumption A.1}

For any $0 < B < \infty$, the semiclassical model has a unique solution of the form
$$
\vec{A}(\vec{x}) = \left\{
\begin{array}{rl}
\frac{B \rho}{2} \vec{e}_{\theta} & \text{if~} \rho \ge R \,\\
a(\rho) \vec{e}_{\theta}          & \text{if~} \rho \le R \,,
\end{array}
\right.
$$
$$\eqno{(2.7)}$$
Equation (2.1) (with $K$ now replaced by the extended region) implies that the state $|\Omega_{\vec{B}})$ is of the form
$$
|\Omega_{\vec{B}}) = \frac{\Psi^{*}(\phi_{0})^{N} |\Omega_{0})}{(N!)^{1/2}}
\eqno{(2.8.1)}
$$
where $|\Omega_{0})$ is the Fock vacuum (no-particle state), and $\phi_{0}$ denotes the normalized ground-state wave function of the one-particle operator
$$
H_{\vec{A}}^{1} \equiv \frac{(\vec{p} - e\vec{A}(\vec{x}))^{2}}{2m}
\eqno{(2.8.2)}
$$
which, in correspondence to the embedding in $\mathbf{R}^{2} \times [-L/2,L/2]$ above, will be considered as an operator on
$$
{\cal H}_{e} \equiv L^{2}(\mathbf{R}^{2},d\mu(\rho,\theta)) \otimes L^{2}((-L/2,L/2))
\eqno{(2.9.1)}
$$
with
$$
d\mu(\rho,\theta) = d\mu(\rho) d\theta \mbox{ with } d\mu(\rho) = \rho d\rho
\eqno{(2.9.2)}
$$
We assume that Neumann boundary conditions are imposed at $z= \pm L/2$. $H_{\vec{A}}^{1}$ may be written
$$
H_{\vec{A}}^{1} = \bigoplus_{k \in \mathbf{Z}} H_{\vec{A}}^{1}(k)
\eqno{(2.10.1)}
$$
where
\begin{eqnarray*}
H_{\vec{A}}^{1}(k) = \frac{\hbar^{2}}{2m}(-\frac{\partial^{2}}{\partial \rho^{2}} - \frac{\partial}{\rho \partial \rho}+\\
+ \frac{|k+ \rho \alpha(\rho)|^{2}}{\rho^{2}} - \frac{\partial^{2}}{\partial z^{2}})
\end{eqnarray*}$$\eqno{(2.10.2)}$$
with $\alpha(\rho)$ given by
$$
\alpha(\rho) = \left\{
\begin{array}{rl}
\frac{e\rho B}{2\hbar}  & \text{if~} \rho \ge R \,,\\
\frac{e a(\rho)}{\hbar} & \text{if~} \rho \le R \,,
\end{array}
\right.
$$
$$\eqno{(2.10.3)}$$
corresponding to the decomposition of the one-particle Hilbert space ${\cal H}_{K} = \oplus_{k \in \mathbf{Z}}{\cal H}_{K}^{(k)}$, with
${\cal H}_{K}^{(k)} = {\cal H}^{(k)} \otimes L^{2}((-L/2,L/2);dz)$ , ${\cal H}^{(k)}$ being the Hilbert space spanned by the orthonormal basis 
$$\{\Psi_{k,j}(\rho,\theta) = (2\pi)^{-1/2} \exp(ik \theta) \Psi_{j}(\rho)\}_{j \in \mathbf{N}} \eqno{(2.10.4)}$$, where $\{\Psi_{j}(\rho)\}_{j \in \mathbf{N}}$ is any orthonormal basis of $L^{2}((0,R);\rho d\rho)$, the latter denoting the Hilbert space of square integrable functions $f(\rho)$ relative to the measure 
$\rho d\rho$ on $\mathbf{R}_{+}$, with the inner product
$$
<f,g> = \int_{0}^{R} \rho \bar{f}(\rho) g(\rho) d\rho
\eqno{(2.10.5)}
$$
$H_{\vec{A}}^{1}(k)$ is a positive self-adjoint operator with purely discrete spectrum and a unique positive g.s. eigenfunction $\phi_{0,e,k}$ under mild conditions on $a(\cdot)$ (see \cite{RSIV}, Theorem X.28, pg. 185 for self-adjointness, \cite{RSIV}, Theorem XIII.47, pg. 207, for discrete spectrum and uniqueness of the g.s.). 

In (2.10.2), (2.10.3), $\rho a(\rho) = f(B)$ is a function of the applied constant field, and (2.10.2) implies that in the g.s. $|k + f(B)|$ must be minimized. Thus $k = k(B)$, and for large $B$ it should be large and negative. For $B$ sufficiently small, however, if $k(\cdot)$ is a continuous function, we have
 $k(B) = 0$. In the latter case, the g.s. wave function $\phi_{0,e}$ of $H_{\vec{A}}^{1}$, given by (2.8.2), satisfies
$$
\partial{\phi_{0,e}(\rho,\theta,z)}{\partial \theta} = 0
\eqno{(2.9.3)}
$$
Unfortunately, the continuity of $k(B)$ seems to be very difficult to prove. Alternatively, since (2.6.2), (2.8.1) and (2.8.2) show that the pair $(\phi_{0,e}, \vec{B})$ is coupled, an assumption only on the smallness of $f(B)$ cannot be made independently of $\phi_{0,e}$. We thus pose

\textbf{Assumption A.2} There exists a $0 < B_{0} < \infty$ such that, if $B < B_{0}$, a solution of the semiclassical problem satisfying (2.9.3) exists.

\textbf{Remark 1} We shall see in lemma 1 that assumption A.2 is equivalent to the assumption of existence of a solution of the semiclassical problem for which only the London form (2.4.2) of the current contributes to the r.h.s. of (2.6.2).

Under assumption A.2, the g.s. of $H_{\vec{A}}^{1}$ lies in the $k=0$ subspace, and the corresponding eigenfunction is
$$
\phi_{0,e,k}(\rho) = \phi_{0,e}(\rho) \frac{1}{L^{1/2}}
\eqno{(2.9.4)}
$$
We shall need a last regularity assumption

\textbf{Asssumption B}
B1)$a(\cdot) \in C^{1}(0,\infty)$ is such that $\vec{B}$, defined by (2.6.3) is continuous and uniformly bounded in $\rho \in [0,\infty)$. Furthermore, we have that: B2) $\phi_{0,e,k}(\rho)$ has a continuous derivative $\phi_{0,e,k}^{'}(\rho)$. 

Concerning assumption B1, we do not expect better than continuity for $\vec{B}$, and boundedness is physically imperative. Assumption B2 is part of the usual assumptions in Sturm-Liouville theory \cite{Weid}. Continuity of $\phi_{0,e,k}^{'}(\rho)$ is used in (A.15) of theorem A.1. The restriction of $\phi_{0,e,k}$ to the cylinder satisfies one of the classical boundary conditions
$$
\frac{\partial f}{\partial n} = \sigma f
$$
where $\frac{\partial f}{\partial n}$ denotes the inward normal derivative, the boundary $\partial K$ being piecewise differentiable in the case of the cylinder (\cite{BRo2}, pg. 55), with $\sigma \in (-\infty, \infty)$. The case $\sigma = \infty$, corresponding to Dirichlet boundary conditions, is excluded by the strict positivity of the g.s. wave-function. It is, however, difficult to say a priori which of these will take place, given that the ''potential'' $a(\cdot)$ is not known explicitly. This is one of the advantages of the extended framework. For the Gaussian, i.e., the uniform induction, $\sigma$ is positive and proportional to $R$. 

A major issue in the Meissner effect and, indeed, in any phenomenon involving currents, is local gauge invariance, our next topic.

\section{Local gauge covariance, its role in the semiclassical model, and properties of the current}

In classical field theory, local gauge invariance or the gauge principle(\cite{MaRo}, Ch. 8, pg. 304, \cite{Thirr}, 4.2, pg. 174) allows one to deduce the form of the field-matter interactions, whereby, specializing to global (space-time independent) transformations, local current conservation is obtained (\cite{Thirr}, 4.2, pg. 174). In the quantum case, local gauge transformations may be defined by
$$
\Psi^{*}(\vec{x}) \to \Psi^{*}(\vec{x}) \exp(ie \alpha(\vec{x})/\hbar)
\eqno{(3.1.1)}
$$
$$
\Psi(\vec{x}) \to \Psi(\vec{x}) \exp(-ie \alpha(\vec{x})/\hbar)
\eqno{(3.1.2)}
$$
$$
\vec{A}(\vec{x}) \to \vec{A}(\vec{x}) + \nabla \alpha(\vec{x})
\eqno{(3.2)}
$$
or, in the smeared form (2.2.2),
$$
U_{\alpha}^{-1} \Psi^{*}(f) U_{\alpha} = \Psi^{*}(\exp(i\alpha)f)
\eqno{(3.3)}
$$
with $U_{\alpha}$ the unitary operator on Fock space given by
$$
U_{\alpha} = \exp(-i \int_{K} d\vec{x} \alpha(\vec{x}) \rho(\vec{x}))
\eqno{(3.4)}
$$
and $\rho$ the density operator
$$
\rho(\vec{x}) = \Psi^{*}(\vec{x}) \Psi(\vec{x})
\eqno{(3.5)}
$$
(this is analogous to \cite{MaRo}, pg. 107). \textbf{Local gauge covariance} is defined in the quantum case by (\cite{Schaf1}, (13.31), pg. 410)
$$
U_{\alpha}^{-1} H(\vec{A}) U_{\alpha} = H(\vec{A} + \nabla \alpha)
\eqno{(3.6)}
$$
which is readily seen (using (2.2.1) to be satisfied by the Hamiltonian $H(\vec{A})$ given by (2.1). Specializing to $\vec{x}$- independent (global) gauge transformations leads to current conservation
$$
\frac{\partial \rho(\vec{x},t)}{\partial t} + \nabla \cdot \vec{j}(\vec{x},t) = 0
\eqno{(3.7)}
$$
with the time dependent operators defined by the Heisenberg picture evolution under $H(\vec{A})$, the operators at zero time being given by (2.3)-(2.4) and (3.5). We have $\frac{\partial \rho(\vec{x},t)}{\partial t} = i [H(\vec{A}),\rho(\vec{x},t)]$, from which
$(\Omega_{\vec{B}},\frac{\partial \rho}{\partial t} \Omega_{\vec{B}}) = 0$ and, thus, by (3.7),
$$
\nabla \cdot (\Omega_{\vec{B}}, \vec{j}(\vec{x}) \Omega_{\vec{B}}) = 0
\eqno{(3.8)}
$$
(3.8) is, of course, a necessary condition for the validity of (2.7), the basic equation of the semiclassical theory. It should be remarked that equations (2.6) should be considered in the distributional sense, i.e., as functionals on $\vec{{\cal D}}(\Gamma)$, the space of vector-valued functions whose components are $C^{\infty}$ functions with compact support in (some open region) $\Gamma$ (see, e.g., \cite{Jo3}, pp 47-51), with the boundary conditions also taken in the distributional sense. This is specially important here, because even in the simple case (2.3)-(2.4), the current is not an operator when taken pointwise, and, thus, (3.7) and (3.8) are formal calculations, which may be made rigorous by employing the smeared forms $\rho \to \rho(f)$, $\vec{j} \to \vec{j}(f)$, where f is a smooth function of compact support in $\mathbf{R}^{3}$; $(\nabla \cdot \vec{j})(f) = - \vec{j} \cdot (\nabla f)$. We shall assume all components of $\vec{f} \in \vec{{\cal D}}(\Gamma)$ equal to a certain $f$. The Maxwell equation (2.6.2) is more precisely stated in the form $(\nabla \times \vec{B})(f) = \mu_{0} (\Omega_{\vec{B}}, \vec{j}(f) \Omega_{\vec{B}})$; taking a sequence 
$\{f_{n}^{\vec{x}}\}_{n \ge 1}$, with $f_{n}^{\vec{x}} \to \delta(\vec{x})$ in the distributional sense, we may obtain (2.6.2) in the form
$$
(\nabla \times \vec{B})(\vec{x}) = \mu_{0} \lim_{n \to \infty} (\Omega_{\vec{B}}, \vec{j}(f_{n}^{\vec{x}}) \Omega_{\vec{B}})
\eqno{(2.6.2)^{'}}
$$
in case the limit on the r.h.s. of $(2.6.2)^{'}$ exists, which will be seen to hold shortly.

We now come back to (2.10.4), and, provisionally, consider $\phi_{0,e}(\rho,\theta)$ instead of $\phi_{0,e,k}$. The restriction of $\phi_{0,e}(\rho,\theta)$ to the cylinder will be denoted by $\phi_{0,e}^{K}(\rho,\theta)$; the variable $z$ will not play any further role beyond (2.9.4) and will be omitted. We have

\textbf{Lemma 1} Assume that $\phi_{0,e}^{K}(\rho,\theta)$ is independent of $\theta$, i.e., satisfies
$$
\frac{\partial \phi_{0,e}^{K}}{\partial \theta} = 0
\eqno{(3.9)}
$$
Then,
\begin{eqnarray*}
\lim_{n \to \infty} (\Omega_{\vec{B}}, \vec{j}(f_{n}^{\vec{x}_{0}}) \Omega_{\vec{B}})=\\
= (\Omega_{\vec{B}}, \vec{j}_{Lon,\vec{B}}(\vec{x}_{0}) \Omega_{\vec{B}})=\\
=-\frac{e^{2}N |\phi_{0,e}^{K}(\rho_{0})|^{2}}{m} a(\rho) \vec{e}_{\theta}
\end{eqnarray*}$$\eqno{(3.10)}$$
for any $\vec{x}_{0}=(\rho_{0},\theta_{0}) \in K$.

\textbf{Proof} We use $(2.6.2)^{'}$ and write
\begin{eqnarray*}
(\Omega_{\vec{B}}, (\Psi^{*} \nabla_{\vec{e}_{\theta}} \Psi) (f_{n}^{\vec{x}_{0}}) \Omega_{\vec{B}}) = \\
= - (\Psi(f_{n}^{\vec{x}_{0}}) \Omega_{\vec{B}}, \Psi(-\nabla_{\vec{e}_{\theta}}f_{n}^{\vec{x}_{0}}) \Omega_{\vec{B}})
\end{eqnarray*}
Using (2.2.3) and (2.8.1),
\begin{eqnarray*}
\Psi(\nabla_{\vec{e}_{\theta}} f_{n}^{\vec{x}_{0}})|\Omega_{\vec{B}}) = \\
=N^{1/2}(\nabla_{\theta}f_{n}^{\vec{x}_{0}},\phi_{0,e}^{K})\\
\frac{\Psi^{*}(\phi_{0,e}^{K})^{N-1}}{((N-1)!)^{1/2}}|\Omega_{0})
\end{eqnarray*}
Since $f_{n}^{\vec{x}_{0}}$ is a smooth approximation of compact support to $\delta(\rho-\rho_{0})(1/\rho)\delta(\theta-\theta_{0})$, an integration by parts, together with (3.9), yields $(\nabla_{\vec{e}_{\theta}}f_{n}^{\vec{x}_{0}}, \phi_{0,e}^{K}) = 0$. Similarly, 
\begin{eqnarray*}
(\Omega_{\vec{B}}, (:\Psi \nabla_{\vec{e}_{\theta}} \Psi^{*}:) (f_{n}^{\vec{x}_{0}}) \Omega_{\vec{B}}) = \\
= (\Omega_{\vec{B}},(\nabla_{\vec{e}_{\theta}}\Psi^{*} \Psi)(f_{n}^{\vec{x}_{0}}) \Omega_{\vec{B}}) = 0
\end{eqnarray*}      
and thus
$$
\lim_{n \to \infty} (\Omega_{\vec{B}}, \vec{j}_{mom}(f_{n}^{\vec{x}_{0}}) \Omega_{\vec{B}}) = 0
\eqno{(3.11)}
$$
A similar, but even simpler, Fock space computation, together with (3.11), yields (3.10). q.e.d.

The standard choice of operators $\Psi$, $\Psi^{*}$, satisfying (2.2.1), implies a choice of phase $\alpha(\vec{x})=0$ in (3.1.1), (3.1.2). Alternatively, from (3.3), this choice corresponds to a choice of a real ground state wave function $\phi_{0}$. Choosing a different $\alpha=\alpha(\vec{x})$ instead, one obtains from (2.4.1) an additional term on the r.h.s. of (3.11),
\begin{eqnarray*}
\lim_{n \to \infty} (\Omega_{\vec{B}}, \vec{j}_{mom}(f_{n}^{\vec{x}_{0}}) \Omega_{\vec{B}})=\\
= \frac{e^{2}N}{m} |\phi_{0,e}^{K}(\rho_{0})|^{2} (\nabla \alpha)(\vec{x}_{0})
\end{eqnarray*}$$\eqno{(3.12.1)}$$
from which
\begin{eqnarray*}
\lim_{n \to \infty} (\Omega_{\vec{B}}, \vec{j}(f_{n}^{\vec{x}_{0}}) \Omega_{\vec{B}})=\\
=-\frac{e^{2}N}{m}|\phi_{0,e}^{K}(\rho_{0})|^{2}(\vec{A}(\vec{x}_{0})- (\nabla \alpha)(\vec{x}_{0}))
\end{eqnarray*}$$\eqno{(3.12.2)}$$
The additional transformation (3.2) yields back the original form (3.10), and, therefore, the result (3.10), under the assumption (3.9), is gauge invariant, although the splitting (2.3)-(2.4) is not. For this reason, (2.4.2) should be written, more precisely:
$$
\vec{j}_{Lon,\vec{B}}(\vec{x}) = -\frac{e^{2}}{m} \Psi^{*}(\vec{x}) \Psi(\vec{x}) (\vec{A}(\vec{x})-(\nabla \alpha)(\vec{x}))
\eqno{(2.4.2)^{'}}
$$
where $\nabla \alpha$ is a gauge function which balances the gauge non-invariance of $\vec{A}$, in order that a gauge invariant combination results.

\section{Main results: the London form of the current and the Meissner effect}

We start our considerations recalling, once again, (2.10.1)-(2.10.4). By the choice (2.7) for the vector potential, which we have shown in section 3 not to entail any loss of generality, only the $\vec{e}_{\theta}$-component of the gradient appears in (2.4.1), which acts on the $\exp(ik \theta)$- part in (2.10.4). From (2.10.2) it is clear that, for $B$ sufficiently large, the ground state will lie in a sector of large (in this case negative) $k$, not in the sector $k=0$ in (2,10.4). Thus, lemma 1 does not apply and both components in (2.3) contribute to the current, and these should cancel each other, being, thus, consistent with $\nabla \times \vec{B} = \vec{0}$ in equation (2.6.2), and a constant $\vec{B} = B \vec{e}_{z}$ penetrating the sample.

Assume, provisionally, that
$$
a(\rho) = \frac{\rho B}{2}
\eqno{(4.1)}
$$
in (2.10.3). When, however, $B$ decreases beyond the value given by
$$
a \equiv \lambda R^{2} < 1/2
\eqno{(4.2)}
$$
with
$$
\lambda = \frac{eB}{2\hbar}
\eqno{(4.3)}
$$
it follows from the simple inequality
\begin{eqnarray*}
\frac{|k+\frac{e B \rho^{2}}{2\hbar}|^{2}}{\rho^{2}} \ge \\
\ge \min_{k \in \{0,-1\}}\frac{|k+\frac{e B \rho^{2}}{2\hbar}|^{2}}{\rho^{2}} \ge\\
\ge (\frac{e B}{2\hbar})^{2} \rho^{2} = \lambda^{2} \rho^{2}
\end{eqnarray*}$$\eqno{(4.4)}$$
that $H_{\vec{A}}^{1}(k)$, given by (2.8.2), satisfies
$$
H_{\vec{A}}^{1}(k) \ge V^{1} - \frac{\hbar^{2}}{2m} \frac{\partial^{2}}{\partial z^{2}}
\eqno{(4.5)}
$$
where
\begin{eqnarray*}
V^{1} = \frac{\hbar^{2}}{2m}(-\frac{\partial^{2}}{\partial \rho^{2}} - \frac{\partial}{\rho \partial \rho}+\\
+ \lambda^{2} \rho^{2})
\end{eqnarray*}$$\eqno{(4.6)}$$
is the harmonic oscillator Hamiltonian in two dimensions. Not unexpectedly, the ground state eigenfunction $\phi_{G}$ is a Gaussian, with
$$
\phi_{G}(\rho,\theta,z)^{2} = \phi_{G}(\rho)^{2} = \frac{h(\rho)}{\pi R^{2} L}
\eqno{(4.7.1)}
$$
with 
$$
h(\rho) = \frac{\lambda R^{2}}{1-\exp(-\lambda R^{2})} \exp(-\lambda \rho^{2})
\eqno{(4.7.2)}
$$
and corresponding eigenvalue
$$
E_{0} = \frac{\hbar^{2} \lambda}{m} 
\eqno{(4.8)}
$$
Thus we see that, under assumption (4.1), (3.9) applies, and the fact that the r.h.s. of (3.10) is nonzero (as long as $\vec{B} \ne \vec{0}$) conflicts with the assumption that $\vec{B} = B \vec{e}_{z}$ inside the sample, which implies $ \nabla \times \vec{B} = \vec{0}$ . This shows that the provisional assumption (4.1) is incorrect, and that the field inside the sample is inhomogeneous. We must therefore come back to the general Ansatz (2.7). For the inhomogeneous problem, the corresponding radial operator is
$$
Q^{1} = \bigoplus_{k \in \mathbf{Z}} Q^{1}(k)
\eqno{(4.10.1)}
$$ 
where
\begin{eqnarray*}
Q^{1}(k) \equiv \frac{\hbar^{2}}{2m} [-\frac{\partial^{2}}{\partial \rho^{2}}\\
 - \frac{\partial}{\rho \partial \rho} + \frac{|k + \rho \alpha(\rho)|^{2}}{\rho^{2}}] 
\end{eqnarray*}$$\eqno{(4.10.2)}$$
$\alpha$ being given by (2.10.3). We also write explicitly the operator corresponding to $k=0$,
 \begin{eqnarray*}
Q^{1}(0) \equiv \frac{\hbar^{2}}{2m} [-\frac{\partial^{2}}{\partial \rho^{2}}\\
 - \frac{\partial}{\rho \partial \rho} +  \alpha(\rho)^{2}] 
\end{eqnarray*}$$\eqno{(4.10.3)}$$
Its ground state eigenfunction $\phi_{0,e}$ may be written in analogy to (4.7) as
$$
\phi_{0,e}(\rho)^{2} = \frac{g(\rho)}{\pi R^{2} L}
\eqno{(4.10.4)}
$$
Note that the normalization (4.10.4) does not mean, of course, that $g$ does not depend on the parameter $R$ - (4.7.2) shows that this is not true in the homogeneous case - but, by assumption B2, we have that
$$
g(\cdot) \mbox{ is continuous in } [0,\infty) \mbox{ and } 0 \le \min_{0\le \rho \le R} g(\rho) \le 1
\eqno{(4.10.5)}
$$
By (2.7), (2.6.3) and assumption B.1,
$$
\frac{\partial (\rho a(\rho))}{\rho \partial \rho} = B(\rho)
\eqno{(4.11.1)}
$$
from which we may choose
$$  
a(\rho) = \frac{\int_{0}^{\rho} d\rho^{'} \rho^{'} B(\rho^{'})}{\rho}
\eqno{(4.11.2)}
$$
We now invoke assumption A.2 and assume that $B < B_{0}$. By lemma 1, together with (4.11.1) and (4.11.2), (2.6.2) leads to the integrodifferential equation
$$
\rho \frac{dB}{d\rho} = \nu g(\rho) \int_{0}^{\rho} d\rho^{'} \rho^{'} B(\rho^{'})
\eqno{(4.11.3)}
$$
and (2.5), to the boundary condition
$$
B(R) = B
\eqno{(4.11.4)}
$$
Above,
$$
\nu = \frac{e^{2} d \mu_{0}}{m}
\eqno{(4.12.1)}
$$
where
$$
d = \frac{N}{V}
\eqno{(4.12.2)}
$$
is the density of particles. The Hamiltonian is $Q^{1}(0)$, given by (4.10.3), and $g(\rho)$ is defined by (4.10.4). 

We write (4.2), (4.3), which were introduced in connection with the homogeneous case, in the form
$$
B(R) = B \in (0,H_{c}(R))
\eqno{(4.13.1)}
$$
with
$$
H_{c}(R) \equiv \frac{\hbar}{eR^{2}}
\eqno{(4.13.2)}
$$

We have 

\textbf{Theorem 1} 
Let
$$
B < \min\{B_{0},H_{c}(R)\}
\eqno{(4.15)}
$$
Then, under assumptions A.1 and B, a necessary and sufficient condion for the existence of the solution specified in assumption A.2 is the inequality
$$
0 \le B(\rho) \le B
\eqno{(4.16.1)}
$$
which, by (4.11.2), is equivalent to the inequality
$$
0 \le a(\rho) \le \frac{\rho B}{2}
\eqno{(4.16.2)}
$$  
where  $B(\cdot)$ denotes the unique bounded solution of (4.11.3), (4.11.4) which follows from assumptions A.1 and A.2. Moreover,
$$
B(\cdot) \mbox{ is monotonically decreasing in the interval } [0,R]
\eqno{(4.16.3)}
$$
Moreover, assumptions A.1 and A.2 are equivalent to the statement that only the London part of the current (2.4.2) contributes to the expectation value on the r.h.s. of the Maxwell equation (2.6.2). In the present context, this expectation value is of the form 
\begin{eqnarray*}
(\Omega_{\vec{B}}, \vec{j}_{Lon,\vec{B}}(\vec{x}) \Omega_{\vec{B}}) =\\
= -\frac{e^{2}}{m} N |\phi_{0,e}(\rho)|^{2} \vec{A}(\vec{x})
\end{eqnarray*}$$\eqno{(4.16.4)}$$
with $\phi_{0,e}$ independent of $\theta$, and
$$
\vec{A}(\vec{x}) = a(\rho)\vec{e}_{\theta}
\eqno{(4.16.5)}
$$
where $a(\cdot)$ is given by (4.11.2), with $B(\cdot)$ the unique solution of (4.11.3), (4.11.4). 

\textbf{Proof} We write (4.11.3) and (4.11.4) together in the form
$$
B(\rho) = B - \nu \int_{\rho}^{R} du g(u)/u \int_{0}^{u} dv v B(v)
\eqno{(4.17)}
$$
or
$$
(AB)(\rho) \equiv  B - \nu \int_{\rho}^{R} du g(u)/u \int_{0}^{u} dv v B(v) = B(\rho)
\eqno{(4.18)}
$$
where  $A$ is an operator from the complete metric space $M_{B,a}$ of continuous functions $h : [R-a,R] \to \mathbf{R}$ such that 
$$
0 \le h(\rho) \le B 
\eqno{(4.19)}
$$
and
$$
||h_{1}-h_{2}|| \equiv \max_{\rho-R \le a}|h_{1}(\rho)-h_{2}(\rho)|
\eqno{(4.20)}
$$
Using, now, (4.10.5), let $L$ be such that
$$
\max_{0 \le \rho \le R} [\nu g(\rho)] \le L 
\eqno{(4.21)}
$$
and $a$ be fixed and chosen such that
$$
a < \min \{1/L, \frac{1}{LB}\}
\eqno{(4.22)}
$$
Then it follows from the definition (4.18), the positivity of $g$, and (4.19) that $A$ maps $M_{B,a}$ into itself and is a contraction from 
$M_{B,a}$ to itself, i.e., 
$$
||A(h_{1}-h_{2})|| < ||h_{1}-h_{2}|| \mbox{ for } h_{1}, h_{2} \in M_{B,a}
\eqno{(4.23)}
$$
Thus, denoting by $B$ the constant function, equal to $B$ for all $\rho \in [0,R]$, (4.17) has a unique solution in the space $M_{B,a}$, which may be uniquely extended to the whole interval $[0,R]$ by the contraction mapping principle (see, e.g., \cite{Arnold}, Theor. 2, Ch. 4, pg. 209). 

Assume, now, that only (4.16.1) and (4.13.1), (4.13.2) hold. The ground state of $Q^{1}$, given by (4.10.1), (4.10.2), lies in the $k=0$ subspace, because
\begin{eqnarray*}
\frac{|k+ \frac{e \rho a(\rho)}{\hbar}|^{2}}{\rho^{2}} \ge \\
\ge \min_{k \in \{0,-1\}}\frac{|k+ \frac{e \rho a(\rho)}{\hbar}|^{2}}{\rho^{2}} \ge\\
\ge \frac{e^{2} a(\rho)^{2}}{\hbar^{2}} 
\end{eqnarray*}$$\eqno{(4.24.1)}$$
because, by (4.16.2) and (4.13.1), (4.13.2), $\sup_{\rho \in [0,R]} \frac{e \rho a(\rho)}{\hbar} < 1/2$. This implies (2.9.3).

The equivalence of (4.16.4) and (4.16.5) to assumption A.2  follows from the fact that, on the one hand, if (2.9.3) is assumed, only the London part of the current (2.4.2) contributes to the r.h.s. of (2.6.2) by lemma 1. The converse follows from the fact that on the r.h.s. of (4.16.4) the wave function must be $\theta$ - independent, a consequence of (2.6.2) and (2.7). Finally, (4.16.3) follows from (4.16.1) and (4.11.3). q. e. d.

\textbf{Corollary 1} Assumptions A.1, A.2 and B imply inequalities (4.16.1) and (4.16.2).

Coming back to the operator $Q^{1}(0)$, given by (4.10.3). Its lowest eigenvalue $\tilde{E}_{0}$ satisfies, by the minimax principle and (4.16.2), 
$$
0 \le \tilde{E}_{0} \le E_{0}
\eqno{(4.24.3)}
$$
where $E_{0}$ is given by (4.8). We note further that $\alpha(\rho)^{2}$ there, given by (2.10.3), satisfies 
$$
0 \le \alpha(\rho)^{2} \le \frac{e^{2} \rho^{2} B^{2}}{4 \hbar^{2}}
\eqno{(4.24.4)}
$$
By (4.10.3), $\alpha(\rho)^{2}$ acts like a positive potential in $\mathbf{R}^{2}$, which is bounded by a harmonic potential. We summarize some known results on $Q^{1}(0)$:

\textbf{Theorem 2} Under assumption B1, $Q^{1}(0)$ is essentially self-adjoint on $C_{0}^{\infty}(\mathbf{R}^{2})$, and has a non-degenerate strictly positive ground state, with wave function $\phi_{0,e}$ satisfying the estimate: for any $a > 0$ , there exists $C_{a} > 0$ such that
$$
|\phi_{0,e}(\rho)| \le C_{a} \exp(-a \rho)
\eqno{(4.25)}
$$

\textbf{Proof} See \cite{RSII}, Theorem X.28, pg. 184; \cite{RSIV}, Theorem XIII.67, pg. 249. (4.26) follows from \cite{RSIV}, Theorem XIII.67, pg. 249, together with the method of proof of \cite{RSIV}, Theorem XIII.70, pg. 251. 

\textbf{Remark 2} It is expected from (2.7) that $\phi_{0,e}$ decays like a Gaussian at infinity.

(4.16.3) allows for an arbitrarily slow decay inside the sample, but the characteristic property of the Meissner effect is the exponential decay of the magnetic induction. For this purpose, a concrete representation of $B(\cdot)$ is required, which is provided by

\textbf{Theorem 3} Under assumptions A.1, A.2 and B, there exists a number $0 < p <\infty$ such that, if
$$
\sqrt{\nu} b > p
\eqno{(4.26)}
$$
then
\begin{eqnarray*}
B(b) \le B \exp[-\frac{\sqrt{\nu}\int_{b}^{R}d\rho (g(\rho))^{1/2}}{2}] \le\\
\le B \exp[-c \sqrt{\nu}(R-b)]
\end{eqnarray*}$$\eqno{(4.27.1)}$$
where $g(\cdot)$ is defined by (4.10.4), and
$$
c = 0.4389
\eqno{(4.27.2)}
$$
The proof of this theorem is given in appendix A (theorems A.1 and A.2). In order to give an idea what is involved in the proof, consider the case
$$
g(\rho) = \mbox{ constant } = s>0
$$
in (4.11.3). Differentiating (4.11.3), we get
$$
\rho \frac{d^{2}B}{d\rho^{2}} + \frac{dB}{d\rho} = \nu s \rho B(\rho)
$$
which, upon division by $\rho$, becomes
$$
\frac{d^{2}B}{d\rho^{2}} +\frac{dB}{\rho d\rho} - \nu s B(\rho) = 0
\eqno{(4.27.3)}
$$
Equation (4.27.3), together with the boundary condition (4.11.4), has the unique bounded solution
$$
B(\rho) = \frac{B I_{0}[(\nu s)^{1/2} \rho]}{I_{0}[(\nu s)^{1/2} R]}
\eqno{(4.27.4)}
$$
where $I_{0}$ is the modified Bessel function of zero order (\cite{ASt}, pg. 374). Since, asymptotically, $I_{0}(z) \sim \exp(z)$ as $z \to \infty$ (\cite{ASt}, pg. 377), (4.27.4) implies that the solution decays exponentially. The proof in appendix A follows this idea closely. The quantity $\nu$, given by (4.12.1), is such that
$$
\sqrt{\nu} = \sqrt{\frac{e^{2}d}{\epsilon_{0} c^{2} m}} = \delta^{-1}
\eqno{(4.27.5)}
$$
Thus,
$$
\delta = \sqrt{\frac{\epsilon_{0} m c^{2}}{d e^{2}}}
\eqno{(4.27.6)}
$$
is the ''penetration depth''. It is independent of the size of the sample and, with physical data, of the order of $1000$ Angstrom, agreeing precisely with the literature (\cite{MaRo}, pg. 165). It is to be noted that, as usual, with macroscopic data, one is within the asymptotic domain (4.26). For example, with $R=10$ cm, $b=R-10 \delta$, say, $b/\delta = 10^{6}$, in the secure domain of asymptoticity of $I_{0}$, with a decay of $\exp(-10)$. Note, however, that these considerations depend on the fact that the constant $s$ in (4.27.1) is a number of order one, or, more precisely (see theorems A.1 and A.2 in appendix A) that the minimum of the function $g(\cdot)$ on $[0,R]$ is of order one. Recalling (4.10.5), we have by theorem 2 that
$$
0 < \min_{\rho \in [0,R]} g(\rho) \le 1
\eqno{(4.27.7)}
$$
The lower bound in (4.27.7) may, however, depend on $R$ (e.g., be $O(1/R)$), or, in principle, be a ridiculously small $R$- independent number. Both alternatives would be inconclusive regarding a physically reasonable ( i.e., of $\delta$ of order (4.27.6)) exponential decay.

The situation is saved, however, by the subharmonic comparison theorem due to Deift, Hunziker, Simon and Vock \cite{DHSV}, which we state in the case of special interest to us:

\textbf{Theorem 4} Let $W \ge V \ge 0$ on $\mathbf{R}^{n}$, $f$ and $g$ be continuous, $\Delta |f| \le V |f|$, and $\Delta |g| \ge W |g|$ in the distributional sense. If $f$ and $g$ $\to 0$ as $x \to \infty$, then $|g| \le |f|$ on $\mathbf{R}^{n}$.

In the case of ground states, $|f| = f$, $|g| = g$ and theorem 4 may be stated as a theorem on the monotony of the g.s. wave-function in the potential, see \cite{Thirr1}, 4.3.32, pg. 232. To sketch the proof for the reader's convenience, let $D = \{x \in \mathbf{R}^{n};g > f \}$. On $D$, 
$\Delta(g-f) \ge Wg-Vf \ge 0$, and thus $g-f$ is subharmonic (see, e.g., \cite{LL}, 9.2, 9.3), and thus attains its maximum on the boundary of $D$, which extends to infinity. Since $f=g \to 0$ as $x \to \infty$, continuity implies that $g$ cannot exceed $f$ on $D$, and therefore $D$ is empty.

Let, now, 
$$
A \equiv Q^{1}(0) - \frac{E_{0}+\tilde{E}_{0}}{2}
\eqno{(4.28.1)}
$$
and
$$
B \equiv V^{1} - \frac{E_{0}+\tilde{E}_{0}}{2}
\eqno{(4.28.2)}
$$
with $V^{1}$ given by (4.6), $Q^{1}(0)$ by (4.10.3), $E_{0}$ by (4.8) and $\tilde{E}_{0}$ the g.s. eigenvalue of $Q^{1}(0)$, satisfying (4.24.3).

\textbf{Theorem 5} Under assumptions A.1, A.2 and B, for $0 \le \rho \le R$,
$$
g(\rho) \ge h(\rho)
\eqno{(4.28.3)}
$$
where $g$ and $h$ are defined by (4.10.4) and (4.7), respectively. Further,
$$
\min_{0 \le \rho \le R} (g(\rho))^{1/2} \ge 0.8779
\eqno{(4.28.4)}
$$

\textbf{Proof} By (4.28.1) and (4.28.2), $\phi_{0,e}$ is the g. s. eigenfunction of $A$, and $\phi_{G}$ the g.s. eigenfunction of $B$, corresponding to the eigenvalues $\frac{E_{0}-\tilde{E}_{0}}{2}$ and $\frac{\tilde{E}_{0}-E_{0}}{2}$, respectively. Thus, identifying $V(\rho) = \alpha^{2}(\rho)$, 
$W(\rho) = \frac{e \rho^{2} B^{2}}{4 \hbar^{2}}$ (see (4.24.4)), we have $0 \le V \le W$ by (4.16.2) and corollary 1. Further, with $f=\phi_{0,e}$, $g=\phi_{G}$, we find $\Delta f - V f \le 0$, $\Delta g - W g \ge 0$. It follows by theorems 2 and 4 that (4.28.3) holds. By (4.8) and (4.2), (4.3), for $a=1/2$, 
$(h(R))^{1/2} = 0.8779 \le (h(\rho))^{1/2} \le (h(0))^{1/2} = 1.1272$, and $(h(\rho)^{1/2}$ grows monotonically when $a$ decreases. This yields (4.28.4). q.e.d.

Employing theorem 5, the reader will find a complete proof of theorem 3 in appendix A (theorems A.1 and A.2).

\subsection{The Meissner effect}

Let
$$
\vec{B}(\rho) = B \vec{e}_{z} \mbox{ for } \rho \ge R 
\eqno{(4.29.1)}
$$
denote, as before, the induction field at the boundary and outside the cylinder. For $B$ satisfying (4.2), (4.3), the magnetic induction inside the cylinder is nonhomogeneous, and therefore the magnetization $M(\vec{x})$ at $T=0$ is defined up to a gradient by the formula of classical magnetostatics
$$
\vec{j}_{Lon,\vec{B}}(\vec{x}) = \nabla \times \vec{M}(\vec{x})
\eqno{(4.29.2)}
$$
The thermodynamic formula for the magnetization 
$$
\vec{M} = -\frac{\partial E_{0}^{N}}{V \partial B} \vec{e}_{z} 
\eqno{(4.29.3)}
$$
where
$$
E_{0}^{N} = N \tilde{E}_{0}
\eqno{(4.29.4)}
$$
is the ground state energy of the system of $N$ particles, requires a uniform source induction field, but should agree, by (4.29.1), with the boundary value of $\vec{M}(\vec{x})$, which we denote by $\vec{M}(R)$. By (4.29.3) and (4.29.4) we thus obtain
$$
\vec{M}(R) = -d \tau \vec{e}_{z}
\eqno{(4.30.1)}
$$
where
$$
\tau \equiv (\frac{\partial \tilde{E}_{0}}{\partial B})_{B=0}
\eqno{(4.30.2)}
$$
By the minimax principle $\tilde{E}_{0}$ grows with $B$, and thus $\tau \ge 0$. We assume
$$
0 < \tau \mbox{ and } \tau \mbox{ is strictly increasing in } B
\eqno{(4.30.3)}
$$
We recall the defining relation for the (applied) magnetic field $\vec{H}$,
$$
\vec{B} = \mu_{0} \vec{M} + \vec{H}
\eqno{(4.31)}
$$
and let
$$
\vec{H} = H \vec{e}_{z}
\eqno{(4.32)}
$$
We also define the quantity
$$
H_{0} = \mu_{0} d \tau
\eqno{(4.33)}
$$

\textbf{Proposition 1 - the Meissner effect} Let $H_{c,R}$ denote the r.h.s. of (4.15). The following assertions hold: a.) if $H_{0} < H <H_{0}+H_{c,R}$ there is exponential decay of the magnetic induction $B$  inside the sample, starting from a boundary value $B_{f} \in (0,H_{c,R})$; b.) $B=0$ for all $H \le H_{0}$; c.)for $H > (H_{0}+H_{c,R})$, $B=H-H_{0}$.

\textbf{Proof} a.) follows directly from theorem 3. For $H=H_{0}$, the boundary value of the magnetic induction is $B=B_{f}=0$ by (4.31). The magnetic induction $B$ remains zero for $H < H_{0}$, because, assuming $B \ne 0$, (4.31) and (4.30.3) yield a negative boundary value $B$, contradicting the assumption (2.5) . This concludes the proof of b.), and c.) follows immediately from (4.31). q.e.d.

Proposition 1 implies that the critical field $H_{c}(0)$, for zero temperature, defined in the description of the Meissner effect given in the introduction, equals
$$
H_{c}(0) = H_{0} + H_{c,R}
\eqno{(4.34)}
$$
Even if $H_{c,R}$ is not optimal, it is expected that $H_{c,R} = o(R)$ as $R \to \infty$, i.e., it should vanish in the thermodynamic limit, being a ''finite-size correction'' to the thermodynamic results ($\vec{H_{0}} = H_{0}\frac{\vec{B}}{|\vec{B}|}$):
$$
\vec{B}(\vec{H}) = 0 \mbox{ if } |\vec{H}| < H_{0} \mbox{ and } = \vec{H} - \vec{H_{0}} \mbox{ if } |\vec{H}| > |\vec{H_{0}}|
\eqno{(4.35)}
$$
obtained by Schafroth (\cite{Schaf1}, pg. 425; \cite{Schaf3}, (5.12), pg. 471). He uses, however, a homogeneous field throughout, and therefore his formulas do not seem to have a bearing to the actual problem. Although assumption (4.30.3) may be argued to be generically true, its verification is nontrivial. The formula for the magnetization implies that the ''active field'' $\vec{H}^{'}$ is identified with the microscopic field, i.e., the magnetic induction $\vec{B}$. This is made plausible by Schafroth (\cite{Schaf3}, pg 471).

\section{Basic features, on the connection to the Higgs effect and open problems}

In this paper we presented a proof, under assumptions A.1, A.2 and B, of the Meissner effect in a (necessarily) gauge covariant model, that of the ground state of free charged Bosons in a constant external magnetic field, in the semiclassical approximation. We summarize here what we found to be the essential basic features underlying the effect, which, in spite of the simplicity of the model, may also be expected to hold in a (still open) more realistic gauge-covariant model with interactions.

Two basic features are: a.) the pairing mechanism: Bosons versus Fermions; b.) finite-size corrections to thermodynamics, surface current and magnetization, the form of the penetration depth and the analogy to charge screening.

Concerning a.), we note that the effect depends crucially on the Bosonic character of the ground state 
$|\Omega_{\vec{B}})$ given by (2.8.1). Although only bosons are considered in this paper, they are interpreted as Schrafoth pairs, i.e., bound states of two fermions, in the applications (see the introduction and the beginning of section 2). We may thus ask what happens with single fermions. For free Fermions in a constant magnetic field, the expectation values of the momentum-density and London parts of the current (2.4.1) and (2.4.2) almost cancel each other, leaving out a weak diamagnetism (see, e.g., \cite{Zi}, pg. 344: a proof proceeds along the same lines by which Landau diamagnetism is proved, see \cite{Zi}, pg. 285). Therefore the ''pairing phenomenon'' in this sense is expected to be crucial.

Point b.) appears in the statement of the Meissner effect, a.) of proposition 1: the effect appears as a finite-size correction to the thermodynamics (4.35). There is an analogy to the charge-screening phenomenon (\cite{MaRo}, pg. 142): (only) an infinite extended system, being an infinite reservoir of particles, accounts for an excess or deficiency of a local charge, not being in contradiction with the conservation of the number of particles. For a finite system, the creation of a polarization cloud around a given charge is accompanied by an accumulation of opposite charges at the surface of the system (surface polarization, see fig. 4.2 of \cite{MaRo}, pg. 143). In close analogy, the estimate (4.27.1) of theorem 3 shows that the current also decays exponentially, and is, thus, also confined to a region of only a few penetration depths from the surface. Therefore, indeed, the essential physical mechanism responsible for the effect are surface currents which create a field $-\vec{B}$ exactly compensating the contribution of the magnetic induction $\vec{B}$ imposed on the sample, at all points inside the metal sufficiently far from the surface (\cite{MaRo}, pg. 166). By (4.29.3), we see that this shielding corresponds to a surface magnetization $\vec{M}$ (analogous to the surface polarization in charge screening), familiar from electrostatics, but obtained here in a quantum context (albeit semiclassical). A very important point is that the penetration depth (4.27.6) is independent of the size of the sample: it agrees, in fact, very well with the conventional theory (\cite{MaRo}, pg. 165).

With regard to the open problems, it is clear that assumptions A.1 and A.2 - the existence of a solution of the London type  - are the most serious and challenging ones. We have already commented on assumption B in the introduction. Concerning the restriction in assumption A.1 to cylindrically symmetric solutions, we should only like to remark that the only symmetry breaking expected in superconductivity (as well as in superfluidity), connected to ODLRO, is the breaking of a global gauge symmetry, see \cite{SeW}, restricted, however, to infinite systems.

Concerning assumption A.1, it should be pointed out that (2.6) is the standard formulation of the ''back-reaction'' of the quantum field $\vec{j}$ on the classical field $\vec{B}$ (see \cite{Schaf1}), analogous to the back reaction of the energy-momentum tensor quantum field upon the space-time geometry in the semiclassical Einstein equation (\cite{Wald}, Chapter 4, pg.98). Under conditions analogous to the ones stated in the latter reference - presently, that the fluctuations of $\vec{j}$ in $\Omega_{\vec{B}}$ are small compared with $|(\Omega_{\vec{B}}, \vec{j} \Omega_{\vec{B}})|$ (which may be expected for macroscopic systems, see, e.g., the analogous discussion of the relative fluctuation of the mean number of particles in the BCS ground state in \cite{MaRo}, pg. 192), and, further, that the fluctuations of the quantized field associated to $\vec{B}$ in the corresponding state (quantum vacuum or thermal state) are small in comparison with the average of the square of the field (\cite{JJS}, pg. 35), the semiclassical model may be expected to be a good approximation. However, as Wald remarks (op. cit. pg. 98), ''even the analogous (to Einstein's equation) justification of the semiclassical Maxwell equation $\nabla^{a}F_{ab} = -4 \pi <j_{b}>$ in quantum electrodynamics has not yet been given''. The only mathematical result we know on the semiclassical model concerns the semiclassical Einstein equation in a cosmological scenario, for which N. Pinamonti \cite{Pina} was able to show the existence of a local solution: for this purpose, the (vacuum) state of the field was assumed to have certain natural properties in that context. It seems, therefore, that the semiclassical model poses an interesting open mathematical problem. Its solution would not only clarify the assumptions A.1, A.2 and B, as well as (4.30.3), but also possibly indicate how to construct a ''self-consistent'' solution starting from initial data, e.g., $(B, \phi_{G}(\rho))$.

We should like to make some brief remarks on the relation to the Higgs effect in relativistic quantum field theory (rqft). This analogy is popularly related to spontaneous symmetry breaking (s.s.b.), see, e.g., \cite{MaRo}, pp. 305-311, but the actual connection between the London current (4.16.4) (with $|\phi_{0,e}(\rho)|^{2} = \frac{\mu_{0}}{V}$ and (4.27.5), (4.27.6), and the identification $(\delta)^{-1} = m$, $m$ being the vector meson mass) and rqft proceeds by way of the so-called Schwinger mechanism of dynamical mass generation \cite{Schw}, see the preface of \cite{FJ} and references given there (Farhi and Jackiw actually refer there to ''London-Schwinger screening''). As remarked in this section, the surface magnetization in the Meissner effect is analogous to the surface polarization in charge screening, and, roughly speaking, the statement analogous to the vanishing of the magnetic induction is that of zero ''charge'' $Q=\int j_{0}(\vec{x})d\vec{x} = 0$, where $j_{0}$ denotes the charge density \cite{Schr}. A deep theorem to this effect in rqft uses local commutativity (microcausality) in an essential and subtle way, and is due to Swieca \cite{Swi} and Buchholz and Fredenhagen \cite{BF}. On the other hand, s.s.b. requires, formally, $Q=\int j_{0}(\vec{x}) d\vec{x} = \infty$, due to existence of vacuum fluctuations occurring over all space, due to translation invariance \cite{SwiecaJ} : these are the real source of s.s.b.. These remarks, due in essence to Schroer (see \cite{Schr} and references given there) exemplify that the usual s.s.b. textbook treatment has no bearing to the real Meissner-rqft analogy. 

The free model treated in this paper could be improved by consideration of the weakly interacting dilute Boson gas of Huang, Yang and Luttinger \cite{HYL} in a magnetic field. In the absence of the field, the model exhibits a gap, which might render it specially interesting as a model of Schafroth pairs in a magnetic field. The present model has been treated from different points of view in \cite{Das}, \cite{FLR}.

\section{Conclusion}

By theorem 1, we have seen that assumption A.2 may be phrased in the equivalent form that only the London part of the current (2.4.2) contributes to the expectation value on the r.h.s. of the Maxwell equation (2.6.2). In most textbooks, and also basic references (\cite{Schaf1},\cite{Schaf3}), it is asserted that the Meissner effect follows, once it is established that the current is of London type; the approach using the macroscopic wave-function \cite{MaRo} also makes this assumption. But - together with assumption A.1 of existence and unicity of a cylindrically symmetric solution, with some regularity properties (assumption B) - this is our basic assumption too! What is, then, the main contribution of the present article?

In our view, our contribution lies in the study of the role played by the seemingly harmless term $|\phi_{0,e}(\rho)|^{2}$ in the expectation value of the London current (4.16.4) (as commented in the previous section, this feature is absent in relativistic quantum field theory, because there the analogue of the square of the wave function is a constant). Since this form is expected to take place only for sufficiently small fields (4.15), the usual treatment of the effect (see \cite{Schaf1} and references given there) uses perturbation theory starting from the free solution, but the latter is not applicable, because, by theorem 2, the spectrum in the presence of any boundary magnetic induction (4.29.1), however small B, is discrete, in contrast to the continuous spectrum in the free case. Most textbooks (see, e,g., \cite{Zi}, pg. 344) assume that the g.s. wave function is, for sufficiently small fields, slowly varying or ''rigid'', i.e., insensitive to the magnetic induction, and thus the g.s. expectation value of the momentum-density part of the current (2.4.1) is ''close'' to the expectation value in the g.s. wave function of the free Hamiltonian, which is zero (see also \cite{Schaf1}).

In theorem 1, it was shown that the assumption that the current is of London type implies the basic inequality (4.16.2). The latter is the main ingredient of the basic theorem 4 on the pointwise monotonicity of the g.s. wave-function on the potential: the homogeneous case provides as comparison wave function the Gaussian solution for the ground state, which is rather flat for small fields. Actually, the wave function inside the sample need not be slowly varying, but, by the bound of theorem 5, it necessarily exceeds the Gaussian wave function pointwise. This is used in theorem 3 to show exponential decay of the magnetic induction inside the sample, with a reasonable value for the penetration depth.

The main difficulties exposed in the previous paragraph remind us that the ground state wave function is basically unknown, because $(\phi_{0,e},\vec{B})$ is a coupled system (see the discussion after (2.10.5) up to remark 1). In having made issues such as this one precise, we hope to  stimulate further mathematical work on this fascinating subject.  

\section{Appendix A - Proof of theorem 3}

We recall (4.11.3),
$$
\rho \frac{dB(\rho)}{d\rho} = \nu g(\rho) \int_{0}^{\rho}du u B(u)
\eqno{(A.1)}
$$
with $\nu$ given by (4.12.1), (4.12.2), and $g$ by (4.10.4). If $g(\rho)$ = constant = $c>0$, we have seen in the main text that the unique bounded solution of (A.1) with the boundary condition
$$
B(R) = B
\eqno{(A.2)}
$$
is given by
$$
B(\rho) = \frac{B I_{0}((\nu c)^{1/2}\rho)}{I_{0}((\nu c)^{1/2} R)}
\eqno{(A.3)}
$$
This motivates the following approach. Consider, now, the operator $A$ defined by (4.18), and choose
$$
0 < \delta < a
\eqno{(A.4)}
$$
where $a$ is given in (4.22), and let $g_{\delta}$ be defined as
$$
g_{\delta} = g(R -l\delta) \chi(R-(l+1)\delta, R-l\delta) \mbox{ for } l=0,1, \cdots, K
\eqno{(A.5)}
$$
where $\chi(a,b)$ is the characteristic function of the interval $(a,b)$, defined as 
$$
\chi(a,b)(\rho) \equiv 1 \mbox{ for } a< \rho \le b \mbox{ and zero otherwise }
\eqno{(A.6)}
$$ 
Let $A_{\delta}$ denote the operator defined as in (4.18), but replacing $g$ by $g_{\delta}$ there. Note that $(A_{\delta}B)$ is a continuous function of $\rho$ for $g \in L^{1}(R-a,R)$ if $B(\cdot)$ is only bounded $L^{1}$. The equation corresponding to (4.17) may be written 
$$
B_{\delta}(\rho) = B - \nu \int_{\rho}^{R} du g_{\delta}(u)/u \int_{0}^{u} dv v B_{\delta}(v)
\eqno{(A.7)}
$$
or, alternatively,
$$
(A_{\delta}B_{\delta})(\rho) = B_{\delta}(\rho)
\eqno{(A.8)}
$$
The proof of Theorem 3 in the main text is completed by the following result:

\textbf{Theorem A.1} Equations (4.18) and (A.7) with the boundary condition (A.2) have both unique solutions in the same space $M_{B,a}$ defined in theorem 1, for any fixed $a$ satisfying (4.22) and any $\delta$ satisfying (A.4), given, respectively, by
$$
B(\rho) = \lim_{m \to \infty} (A^{m} B)(\rho)
\eqno{(A.9)}
$$
and
$$
B_{\delta}(\rho) = \lim_{m \to \infty} (A_{\delta}^{m} B)(\rho)
\eqno{(A.10)}
$$
Moreover, the above solutions may be uniquely extended to the whole interval $[0,R]$. Denoting the extensions by the same symbols, there exists a constant $f > 0$ such that
$$
\max_{0 \le \rho \le R} |B(\rho) - B_{\delta}(\rho)| \le f \delta
\eqno{(A.11)}
$$
In (A.9), (A.10), $B$ denotes the constant function, equal to $B$ for all $\rho \in [0,R]$, and $A^{m}$ denotes the $m-$ fold composition of $A$ with itself.

\textbf{Proof} Since $|g_{\delta}| \le L$ with the same $L$ as $g$ by the definition (A.5), (A.6), and $g_{\delta} \ge 0$, $A_{\delta}$ is also a contraction from $M_{B,a}$ to itself. The first assertion of theorem A.1, together with (A.9) and (A.10), follow from the contraction mapping principle (see, e.g., \cite{Arnold}, Th. 2, Ch. 4, pg. 209). The existence of unique extensions to the whole interval follows from the same theorem.

We now prove (A.11). Because $A$, defined by (4.18), and $A_{\delta}$, defined by (A.8), are not linear operators, due to the $B$- term in (4.17) (resp. (A.7)), we express equations (4.17) (resp. (A.7)) in terms of the operators $C$ (resp. $C_{\delta}$) obtained by omitting the term $B$ in (4.17) (resp. (A.7)). The operators $C$ and $C_{\delta}$ are then linear, their contractivity follows from that of $A$ and $A_{\delta}$, and
$$
C^{m} - C_{\delta}^{m} = \sum_{\gamma=1}^{m} C^{m-\gamma} (C-C_{\delta}) C_{\delta}^{\gamma-1}
\eqno{(A.12)}
$$
(A.12) is valid for any two bounded linear operators, such as $C$ and $C_{\delta}$ on $M_{B,a}$.  We now apply both members of (A.12) to $B$ Since $C$ and 
$C_{\delta}$ are contractions from $M_{B,a}$  to itself,  
$$
||(C^{m} - C_{\delta}^{m})B|| \le \sum_{\gamma=1}^{m} ||C^{m-\gamma}|| ||C-C_{\delta}|||C_{\delta}^{\gamma-1} B||
\eqno{(A.13)}
$$
where $||A||= \sup_{f \in M_{B,a}} ||Af||/||f||$ is the operator norm on $M_{B,a}$. By definitions of $C$, $C_{\delta}$, $g$ and $g_{\delta}$, we obtain
$$
||C-C_{\delta}|| \le \nu B r \delta
\eqno{(A.14)}
$$
where
$$
r = \max_{\rho\in[0,R]} |g^{'}(\rho)|
\eqno{A.15)}
$$
which is finite by assumption B2. Representing now the functions $B(\rho)$ and $B_{\delta}(\rho)$ as $M_{B,a}$ vectors $B$ and $B_{\delta}$, eqs. (4.17) and (A,7) are equivalent to $(\mathbf{1} + C)B = B(R)$ and $(\mathbf{1} + C_{\delta}) B_{\delta} = B(R)$, respectively. By the invertibility of $\mathbf{1} + C$ and $\mathbf{1} + C_{\delta}$, their solutions are expressible in terms of power series in $C$ and $C_{\delta}$, we obtain (A.11) from the estimates (A.13) and (A.14, with the interval replaced by $[R-a,R]$. By the aforementioned extension, the result follows in general. q.e.d.

As indicated in the main text, the solution of (A.7) is of the form
$$
\alpha_{0} I_{0}[(\nu c_{0})^{1/2} \rho] \mbox{ with } c_{0} = (\nu g(R))^{1/2}
\eqno{(A.16)}
$$
with $\alpha_{0}$ determined by the boundary condition (A.2),
$$
\alpha_{0} = \frac{B}{I_{0}[(\nu c_{0})^{1/2} R]}
\eqno{(A.17)}
$$
In the subsequent intervals $[R-(i-1) \delta, R-i \delta]$, $i=1,2, \cdots, ,N$, the solution is determined by continuity at the upper boundaries and equals
 
$$
B_{\delta}(R-i\delta)=B \prod_{j=0}^{i-1} \frac{I_{0}[(\nu c_{j})^{1/2} (R-(j+1)\delta)]}{I_{0}[(\nu c_{j})^{1/2} (R-j\delta)]}
\eqno{(A.18)}
$$
with
$$
c_{j} = g(R-j \delta)
\eqno{(A.19)}
$$
Let, now,
$$
R-N \delta = b
\eqno{(A.20)}
$$
Theorem 3 of the main text now follows from

\textbf{Theorem A.2} Let assumptions A.1, A.2 and B hold and let $B(\cdot)$ denote the unique solution of (4.17). Then there exists a $0 < p < \infty$ such that, if
$$
(\nu)^{1/2} b > p
\eqno{(A.21)}
$$
then
\begin{eqnarray*}
|B(b)| \le B \exp[-\frac{(\nu)^{1/2} \int_{b}^{R} d\rho (g(\rho))^{1/2}}{2}]\\
\le B \exp[-c (\nu)^{1/2} (R-b)]
\end{eqnarray*}$$\eqno{(A.22.1)}$$
with
$$
c = \frac{0.8779}{2} = 0.4389
\eqno{A.22.2)}
$$

\textbf{Proof} We write in (A.18)
$$
I_{0}[(\nu c_{j})^{1/2} (R-(j+1)\delta)] = I_{0}[(\nu c_{j})^{1/2} (R-j\delta)] - I_{1}(\lambda_{j})(\nu c_{j})^{1/2} \delta
$$
where
$$
\lambda_{j} \in ((\nu c_{j})^{1/2} (R-(j+1)\delta),(\nu c_{j})^{1/2} (R-j\delta))
$$
Thus,
\begin{eqnarray*}
\frac{I_{0}[(\nu c_{j})^{1/2} (R-(j+1)\delta)}{I_{0}[(\nu c_{j})^{1/2} (R-j\delta)}=\\
=1- \frac{I_{1}(\lambda_{j})(\nu c_{j})^{1/2} \delta}{I_{0}[(\nu c_{j})^{1/2} (R-j\delta)]}\\
\le \exp[-\frac{I_{1}(\lambda_{j})(\nu c_{j})^{1/2} \delta}{I_{0}[(\nu c_{j})^{1/2} (R-j\delta)]}]
\end{eqnarray*}
by the inequality $1-x \le \exp(-x)$, true if $x \ge 0$. Inserting this estimate in (A.18), we get
$$
B_{\delta}(b) \le B \exp[-\sum_{j=0}^{N-1}\frac{I_{1}(\lambda_{j})(\nu c_{j})^{1/2} \delta}{I_{0}[(\nu c_{j})^{1/2} (R-j\delta)]}] 
\eqno{(A.23)}
$$
We now use (A.11) and the fact that we have, by (A.19), a Riemann sum for the function 
$\frac{I_{1}((\nu g(\rho))^{1/2}\rho)}{I_{0}((\nu g(\rho))^{1/2}\rho)}(\nu g(\rho))^{1/2}$ inside the exponential in (A.23), to obtain
$$
|B(b)| \le B \exp[-\int_{b}^{R} d\rho \frac{I_{1}((\nu g(\rho))^{1/2}\rho)}{I_{0}((\nu g(\rho))^{1/2}\rho)} (\nu g(\rho))^{1/2}]
\eqno{(A.24)}
$$
The asymptotic expansions for $I_{0}$ and $I_{1}$ (\cite{ASt}, pg. 377) now yield that, for given $0 < p < \infty$, 
$$\frac{I_{1}((\nu g(\rho))^{1/2}\rho)}{I_{0}((\nu g(\rho))^{1/2}\rho)} \ge 0.5 \eqno{(A.25)}$$ if $b$ satisfies (A.21). Inserting (A.25) and (4.28.2) of theorem 5 for the term $(\nu g(\rho))^{1/2}$ into (A.24), we obtain the last bound in (A.22.1), with $c$ satisfying (A.22.2). q.e.d.

\textbf{Acknowledgement} We should like to thank G. L. Sewell for his interest in this work and for critical remarks on the first version of these notes, and, most specially, to the referee, for numerous suggestions, including a complete revision of the manuscript, for pointing out an error in the previous theorem A.1 and suggesting how it could be corrected, and insisting on the necessity of assumption A.2, which is crucial.

\bibliography{minhabiblio}
\bibliographystyle{alpha}
\end{document}